\documentclass[12pt,manuscript]{aastex}
\usepackage{emulateapj5}
\usepackage{psfig}

\begin{document}

\newcommand{\casa}{Cas~A}
\newcommand{\Te}{$kT_{\rm e}$}
\newcommand{\net}{$n_{\rm e}t$}
\newcommand{\netunit}{${\rm cm}^{-3}{\rm s}$}
\newcommand{\EM}{${n_{\rm H} n_{\rm e} V}/{4\pi d^2}$}
\newcommand{\NH}{$N_{\rm H}$}
\newcommand{\NHunit}{${\rm cm}^{-2}$}
\newcommand{\fluxunit}{${\rm ph cm^{-2} s^{-1}}$}
\newcommand{\msun}{M$_{\sun}$}
\newcommand{\spex}{{\it SPEX}}
\newcommand{\xspec}{{\it xspec}}
\newcommand{\ciao}{{\it Ciao}}
\newcommand{\vmekal}{{\it vmekal}}
\newcommand{\sresc}{{\it sresc}}
\newcommand{\srcut}{{\it srcut}}
\newcommand{\tiff}{$^{44}$Ti}
\newcommand{\caff}{$^{44}$Ca}
\newcommand{\scff}{$^{44}$Sc}
\newcommand{\jetp}{JETP}
\newcommand{\adspr}{AdSpR}
\newcommand{\phrvl}{PhRvL}
\newcommand{\phrc}{PhRvC}
\newcommand{\rvmp}{Rev. Mod. Physics}
\newcommand{\xmm}{{\it XMM-Newton}}
\newcommand{\chandra}{{\it Chandra}}
\newcommand{\asca}{{\it ASCA}}
\newcommand{\cangeroo}{{\it CANGEROO}}
\newcommand{\hegra}{{\it HEGRA}}
\newcommand{\sax}{{\it BeppoSAX}}
\newcommand{\osse}{{\it CGRO}-OSSE}
\newcommand{\egret}{{\it CGRO-EGRET}}
\newcommand{\integral}{{\it INTEGRAL}}
\newcommand{\glast}{{\it GLAST}}
\newcommand{\comptel}{{\it CGRO-COMPTEL}}
\newcommand{\iras}{{\it IRAS}}

\title{On the magnetic fields and particle acceleration in Cas A}
\shortauthors{J. Vink \& J. M. Laming}

\author{Jacco Vink\altaffilmark{1}}
\affil{Columbia Astrophysics Laboratory, Columbia University, MC 5247, 
550 W 120th street, New York, NY 10027, USA}
\email {jvink@astro.columbia.edu}
\altaffiltext{1}{Chandra fellow}

\author{J. Martin Laming}
\affil{Naval Research Laboratory, Code 7674L, Washington DC 20375, USA}
\email{jlaming@ssd5.nrl.navy.mil}

\begin{abstract}
We investigate the non-thermal X-ray emission from \casa,
using \sax, \osse, and \chandra\ data. For the hard X-ray continuum we
test the model proposed by Laming, which invokes
non-thermal bremsstrahlung from electrons accelerated by lower hybrid plasma 
waves.
The justification for this model comes from our determination of a lower limit
to the average magnetic field of $B > 0.5$~mG.
For such high magnetic fields the synchrotron
losses are severe enough that most of the electron populations responsible
for the radio emission have maximum electron energies well below the limit for
which X-ray synchrotron emission is important.
However, we do suggest that the rim surrounding \casa, 
seen in \chandra\ continuum images, is X-ray synchrotron emission.
The width of this rim of 1.5\arcsec\ to 4\arcsec, can be used to infer the
magnetic field near the shock front, for which we estimate $B = 0.08 - 0.16$~mG,
and electron energies of $\sim 57 - 40$~TeV.
This magnetic field strength  is lower than the average magnetic field, but
higher than what may be expected from shocked interstellar medium,
suggesting either a high magnetic field in the wind of the progenitor,
or rapid, post shock, field amplification by non-linear growth of plasma waves.
Combining the two magnetic field measurements we have constructed a
simple two zone model. 
Most of the radio emission comes from inside \casa\, where
the magnetic field is strong. 
In contrast, the inverse Compton emission is dominated by
emission from near the shock front. 
Only for our lower limit on the magnetic field
strength near the shock front is it possible to explain the recent 
detection of TeV emission by \hegra\ with inverse Compton, for which, 
in addition, we have to assume
a rather high far infrared photon density that should be twice as high as
our best estimate of $\sim 70$~cm$^{-3}$. Pion decay is therefore likely to be
the dominant emission from \casa\ at TeV energies.
\end{abstract}


\section{Introduction}
Supernova remnants have long been thought to be the source of cosmic rays 
with energies up to at least $10^{15}$~eV, 
where there is a break in the observed cosmic ray spectrum, usually referred
to as the ``knee''. 
The recent discoveries of X-ray synchrotron and TeV emission from some
supernova remnants support this long held belief by confirming that
electrons are accelerated to energies of at least a few TeV \citep{Reynolds99}.
This is still somewhat short of the ``knee'', but unlike for ions, the maximum
electron energy may be limited by radiation losses.

SN 1006  has become the canonical example of
a supernova remnant with X-ray emission dominated by
synchrotron radiation from shock accelerated electrons \citep{Koyama95}.
The presence of electrons in excess of 10~TeV is further supported by the detection of 
TeV emission with the \cangeroo\ \v Cerenkov detector \citep{Tanimori97}. 
The observed TeV emission is most likely inverse Compton 
scattering by the same population of electrons that is responsible for the 
X-ray continuum \citep{Tanimori97},
although there is some debate about this interpretation \citep{Berezhko02}.

Another often mentioned candidate for an X-ray synchrotron emitting supernova 
remnant is \casa. 
Its soft X-ray emission is not continuum dominated,
but the observed emission above 20~keV
cannot be explained by thermal bremsstrahlung, and has therefore
been attributed to synchrotron radiation \citep{The96, Allen97,Favata97}.
The detection of TeV emission by \hegra\ 
is consistent with this
interpretation, but pion decay associated with relativistic ions is
an equally valid interpretation \citep{Aharonian01}.

For \casa\ the interpretation of both the hard X-ray and the
TeV emission is critically dependent on the magnetic field strength.
Equipartition arguments to explain the radio luminosity 
imply magnetic fields of 
$B \sim$ 0.4 - 2 mG \citep{LongairV2,Wright99}, 
far in excess of the canonical value for the interstellar
medium of $3\, \mu$G.
The typical loss time of an electron with energy $E$,
due to synchrotron radiation is \citep[e.g.][]{Reynolds98}:
\begin{equation}
\tau_{loss} = \frac{635}{B^2 E}\, {\rm s}.\label{eq-losses}
\end{equation}
For a magnetic field of 1~mG, a particle with $E = 8$~TeV 
loses its energy in 2~yr. For \casa, 8~TeV is
the maximum energy needed to explain the hard X-ray emission 
\citep{Reynolds99}.
The two year lifetime is much shorter than the lifetime of \casa, which
is probably the remnant of a supernova that exploded in AD 1680 
\citep{Ashworth80}. 
Similarly a high magnetic field lowers the number density of 
relativistic electrons for a given radio brightness, 
which in turn implies a low inverse Compton contribution to the
observed TeV emission.

One way out of this situation would be to
assume that electrons are continuously being reaccelerated,
but that seems unlikely. The radio morphology of \casa\ shows
a bright ring of radius $\sim$1.8\arcmin, whereas the shock front,
the primary site for cosmic ray acceleration,
is at a radius of $\sim$2.6\arcmin.
Moreover, detailed radio spectral index studies do not find evidence
of reacceleration in bright radio knots, which are probably
ejecta fragment penetrating the shell \citep{Anderson96,Wright99}.
These studies rather suggest that brightness variations are
the result of variations in the magnetic field, which light up the
diffuse relativistic electron population in various degrees.

Here we test an alternative interpretation of the
hard X-ray emission with \sax-PDS data,
namely bremsstrahlung from electrons accelerated
by lower hybrid waves, which was suggested by 
\citet{Laming01a,Laming01b}.
An additional argument for testing such a model is that \xmm\ observations
show that even above 10~keV a substantial fraction of the continuum emission
is not associated with the forward shock \citep{Bleeker01}.
The \sax\ data, in combination with \osse\ data,
also provides an upper limit on the 
bremsstrahlung from the electron cosmic ray injection spectrum,
from which a new lower limit on the 
average magnetic field strength can be obtained.

Nevertheless, we argue that X-ray synchrotron radiation does exist in \casa,
and is coming from close to the forward shock, where it shows up
as a narrow  rim surrounding the remnant, observed in \chandra\ X-ray 
continuum images \citep{Gotthelf01}.
Interpreting the width of the rim as a typical length scale for
synchrotron losses, we are able to constrain the magnetic field near
the shock front.
Our analysis supports the idea that the average magnetic field is
close to its equipartition value,
whereas we find that near the shock front the magnetic field is lower,
but still an order of magnitude higher than what may be expected
from shocked interstellar medium ($\sim 10^{-5}$G).
We will discuss the implications of these values for the hard X-ray
and TeV emission.

\section{Data and observations}
Our analysis of the X-ray emission of \casa\ is based on
\sax-PDS data and archival \osse\, and \chandra\ data.
The \sax-PDS \citep{Frontera97} and \osse\ \citep{Johnson93}
instruments both consist of four phoswitch detectors
(NaI(Tl)/CsI(Na) scintillation detectors), behind 
two rocking collimators. They differ, however, substantially
in energy range, field of view, and effective area. 
The \sax-PDS instruments cover an energy range of 15-300~keV,
whereas this is 50~keV  to 10~MeV for \osse.

\sax\ observed \casa\ for 500~ks in May and June 2001 in order
to observe the 67.9 ~keV and 78.4~keV nuclear decay
lines of \scff, the nuclear decay product of \tiff.
The detection of these lines at the 3-5$\sigma$ level, depending on
the hard X-ray continuum model, was reported by \cite{Vink01},
which provides more details about the data set.
Due to the rocking collimators the effective exposure time is about half the 
observation time. Together with archival \sax\ data the spectra used
for our analysis here is based on a total effective PDS exposure of 311~ks.

The spectrum, as presented in \citet{Vink01}, 
clearly shows the hard X-ray spectrum with a bump associated with the 
\scff\ line emission.  Marginal (2$\sigma$) residual emission around 60~keV 
is possibly caused by fluorescent tantalum K shell emission, as a result of 
cosmic rays interacting with the collimator, which consists of tantalum with 
a thin copper-tin bi-layer.
Some residual emission above 100~keV is not easily explained 
(cf. Fig.~\ref{fig-pds} \& \ref{fig-osse}). 
It may be intrinsic to \casa, 
but some instrumental contamination seems more likely. 
The residual emission peaks around 200~keV, but
is not narrow enough to be caused by fluorescent line emission from any
particular material.
There are no other similar deep \sax\ observations with a moderate hard X-ray
brightness like \casa, which would allow us to identify possible instrumental 
features at this level.

Although the residual emission above 100~keV is statistically not very 
significant,
one of our concerns is to obtain an upper limit on the emission
above 100~keV in order to obtain a lower limit to 
the average magnetic field in \casa.
We therefore included archival \osse\ spectra in our analysis.
Moreover, the effective area of the instruments peaks around 250~keV, 
making it useful for constraining the emission of \casa\ above 100~keV.
\casa\ was observed several time by \osse\ \citep[see][]{The96}. 
We added all available \casa\ spectra 
and averaged the corresponding response matrices. 
The observations were made during the viewing periods
34 - 815 with a total exposure of 909~ks.\footnote{In addition to the observations listed in 
\citet{The96}, this means VP 617, 715, 810, 814, and 815.}
The total exposure time was 909~ks.

In addition to hard X-ray data, we use a deep (50 ks) archival \chandra\ observation
with the ACIS-S3 chip to complement our analysis. The observation was
made in January 2000. The same data were used by \citet{Hwang00}.
For our analysis we used level 2 archival data products, 
generating spectra and calibration products with 
the standard \chandra\ software package \ciao\ v1.1.0.

\section{The lower hybrid electron acceleration model}
\label{sec-lhw} 
\casa's morphology indicates a complicated,
turbulent hydrodynamical structure. Radio, optical and X-ray
images all show many bright knots on top of more diffuse emission.
As the forward or reverse shocks strike and illuminate these
features, secondary shocks will split off from these primary shocks
and propagate back into the already shocked circumstellar or
ejecta plasma. In this way shocks may propagate across the likely
region of high magnetic field at the contact discontinuity, where 
field amplification by Rayleigh-Taylor instability may occur. The
acceleration of electrons by lower hybrid waves generated by these
shocks has been considered by \citet{Laming01a,Laming01b}. Bremsstrahlung
emitted by these accelerated electrons was shown to provide a very
good match to the hard X-ray continuum of \casa, using the $\sim$
50~ksec of \sax-PDS data that were available at the time.

Lower hybrid waves (LHW) are electrostatic ion oscillations directed
within an angular range $\pm\omega _{pi}/\omega _{pe}$ of the direction
perpendicular to the magnetic field, where $\omega _{pe}$ and
$\omega _{pi}$ are the electron and ion plasma frequencies
respectively. The electron screening that would usually damp such
oscillations is inhibited by the magnetic field, giving a
criterion on the electron temperature that the electron gyroradius
be smaller than the wavelength divided by $2\pi $. Since
$\omega /k_{\perp} <<\omega /k_{||}$ the wave can simultaneously
be in resonance with ions moving across the magnetic field lines
and electrons moving along them. Consequently energy equilibration
between ions and electrons may proceed at a much faster rate than
by Coulomb collisions alone, and saturate when the electron
thermal pressure is equal to the magnetic pressure. Such processes
have been considered in the context of two-temperature accretion
flows by \citet{Begelman88}. Here cross field ion motions
are produced by a combination of curvature and diamagnetic drift,
and the magnetic field is presumably generated by a
magneto-rotational instability \citep{Balbus91,Balbus98,Quataert02}, 
the critical issue being whether sufficient
electron-ion equilibration can occur to destroy ADAF or similar
solutions for the accretion flow. In dilute plasmas where the
electron acceleration time is much shorter than the
electron-electron collision time, a non-Maxwellian electron
distribution in the component of velocity along the magnetic field
can result. This has been modeled numerically using
particle-in-cell simulations by \citet{Shapiro99} for
application to cometary X-rays, and analytically by \citet{Vaisberg83}
and \citet{Kranoselskikh85}. Lower hybrid waves
generated by a modified two-stream instability and the resulting
accelerated electrons were all observed at Comet Halley
\citep{Klimov86,Gringauz86}. \citet{Bingham00} considered
LHW in supernova remnants and \citet{Laming01a,Laming01b} applied the analytic
formulations, which agree well with the numerical work, to the
case of accelerated electrons in \casa. The non-Maxwellian
electron distribution function depends on two parameters, the
overall normalization (i.e. the fraction of plasma electrons that
are accelerated) and the maximum electron energy in the
distribution. In our fitting we treat these both as free
parameters. A demonstration of how  these variables are related to
other physics of the \casa\ plasma, and hence could be determined
is given in the Appendix.

We fitted the predicted spectrum, calculated as in \citet{Laming01a},
but with a relativistic bremsstrahlung cross section taken from
\citet{Nozawa98},  to the observed \sax-PDS
X-ray spectrum. Apart from the  non-thermal bremsstrahlung spectrum several
other components, which are likely to contribute to the hard X-ray emission,
are included in the spectral model.
At the low end of the hard X-ray spectrum, the tail of the
the thermal bremsstrahlung with $T_{\rm e} = 4\times10^7$~K is expected to
contribute to the spectrum.
In addition, \scff\ line emission at 68 and 78 keV is expected,
as a result of the decay of \tiff\ synthesized during the explosion
\citep{Vink01,Iyudin94}.
The line fluxes of the two nuclear transitions are expected to be almost identical.
A broadening corresponding with a velocity dispersion of 5000~km~s$^{-1}$\ was
included. A residual feature around 60~keV in the observed spectrum may be
attributed to cosmic ray induced fluorescence from the tantalum collimator,
and was modeled by delta functions corresponding to
Ta K$\alpha$\ (57~keV) and K$\beta$ (65~keV) with a ratio of 4:1.\footnote{
See the X-ray data booklet \citep{xraybooklet}.}
The spectrum and best fit model are shown in Fig.~\ref{fig-pds} and
model parameters
are given in Table~\ref{tbl-lhw}.

\medskip
\vbox{
\centerline{\psfig{figure=casa_lhw_fig.ps,width=8cm,angle=-90}}
\figcaption{\sax-PDS spectrum of Cas A with the best fit \scff/non-thermal 
bremsstrahlung model. 
The individual emission components are:
\scff\ line emission (red), 
non-thermal continuum (blue), thermal continuum
(green) and possible line contamination from collimator
material (tantalum, cyan). The observed count rate in each channel 
has been divided by the effective area in order to yield approximate flux 
densities.\label{fig-pds}}}
\bigskip

The parameter values indicate that electrons are possibly
accelerated up to $\sim$95~keV.
The model gives a very good fit to the spectrum up to 100~keV.
However, it does not describe
the emission observed above 100~keV.
As discussed above, the data above 100~keV may be less reliable,
and the low flux point around 100~keV suggest that the actual continuum emission
is indeed curved, as predicted by the LHW model.

In Table~\ref{tbl-lhw} we also list for comparison
the parameters for a simple power law continuum spectrum,
and the X-ray synchrotron models available
in the spectral fitting package \xspec\ \citep{xspec},
\srcut\ and \sresc\ \citep{Reynolds98,Reynolds99}.
The \srcut\ model is a very simplified model of X-ray synchrotron emission
from a relativistic electron distribution with an exponential cut-off.
In reality the cut-off energy is likely to vary from
place to place, giving a more gradual decrease in slope.
Moreover, as explained in the introduction, for the bulk of the
relativistic electrons, at a remote distance from the shock front,
the maximum electron energy is probably far below
what is needed to produce X-ray synchrotron emission.
The \sresc\ model takes these spatial effects into account,
assuming a Sedov model for the evolution
of the supernova remnant. Although, \sresc\ is more physical, it may not
be readily applicable to \casa, which has large
magnetic field gradients and is not well described by the standard
Sedov model \citep{Vink98}.

The LHW model is in a similar way somewhat oversimplified,
as also here it is likely that maximum energies
vary within the remnant,
depending on local conditions such as the
local shock velocity, obliquity and ion composition.
Nevertheless, given the limited number of parameters,
the model fits very well.
Note that the spectral slope between 20-50 keV, is a property of the model,
and not a free parameter.

Finally, note that using the LHW model
results in 50\% higher \scff\ line fluxes, than using a power law continuum,
a result of the continuum curvature predicted by both models.
The higher \scff\ line flux is close to the latest \comptel\ measurements of
the \caff\ line flux \citep{Schoenfelder00},
and translates into a synthesized \tiff\ mass of
$1.8 \times 10^{-4}$~\msun, for a distance of 3.4~kpc \citep{Reed95} and
a \tiff\ decay time of 85.4~yr \citep{Ahmad98,Goerres98,Norman98}.

\section{A new lower limit to the average magnetic field}
The LHW model is attractive to consider for \casa\
because of its high magnetic field strength,
which is estimated to be $\sim$1~mG, based on the assumption
of equipartition of the relativistic particle population and the magnetic 
field energy \citep[e.g.][]{LongairV2}.
Although there is no compelling physical reason why 
the equipartition assumption should be valid,
estimates based on the absence of gamma-ray emission from \casa\ supports
a value of the magnetic field in excess of what might be expected from a shock
compression of the interstellar magnetic field \citep{Cowsik80}. 
The most recent
estimate of the magnetic field is based on the upper limit obtained by \egret,
suggesting $B > 3.5 \times 10^{-4}$~G \citep{Esposito96,Atoyan00}.
This value is derived by comparing the radio synchrotron flux density with the
expected bremsstrahlung from the relativistic electrons responsible for the 
radio emission.

Here we use essentially the same method, but less directly, 
as we use an extrapolation
of the relativistic electron population to energies observed by \sax-PDS 
and \osse.
The extrapolation used is similar to the one by \citet{Asvarov90}, 
and assumes that
the electron {\em momentum} spectrum is a power law distribution, 
as is expected for an electron population generated by
first order Fermi acceleration \citep{Bell78}.
This means that the {\em energy} spectrum flattens at non-relativistic 
energies.
The electron cosmic ray injection spectrum is not well known, 
but models indicate that at lower energies the spectrum is steeper than 
expected from extrapolating the
relativistic spectrum \citep{Bykov99}.
In that case our upper limit on the cosmic ray normalization is too high,
which translates into too low a lower limit.

Of more concern is the role of energy losses on the shape of the spectrum, 
as it will tend to
flatten the low energy end of the spectrum.
We checked the signature of adiabatic losses on the spectrum, and found that 
it does affect the spectrum mostly at mildly relativistic energies, 
where it results in a steeper spectrum at low energies, is flatter at 
intermediate energies
and then assumes the original power law index at extreme relativistic energies.
We did not take the steepening into account, as we do not know how much the 
average electron spectrum is affected by it. 
But if there is a steepening, it makes our estimates too conservative.
The most dominant other loss factor between 100 and 1000 keV in the 
ionized plasma of the remnant is electron-electron collisions.
For an electron temperature of $4\times10^7$~K and density of 20~cm$^{-3}$, 
the loss time of a 100~keV electron is $\sim$200~yr, 
rapidly increasing to $\sim$500 yr for 200~keV \citep{nrlbooklet}.
These loss times are comparable to or larger than the average plasma age 
of \casa, indicating that non-adiabatic losses are likely to be small.

The synchrotron flux density for a relativistic electron power law 
distribution, with energy index $q$, is given by \citep{Blumenthal70}:
\begin{equation}
F_{\nu} = 1.70\times 10^{-21} a(q) \frac{\kappa V}{4\pi d^2} B^{\frac{q+1}{2}} 
\bigr(\frac{6.26\ 10^{18}}{\nu})^{\frac{q-1}{2}}\label{eq-synchrotron}
\end{equation}
where $\kappa$\ is the particle normalization, $V$ the emitting volume, 
$d$ the distance, $B$ the magnetic field, 
and $a(q)$ is a factor tabulated in \citet{Blumenthal70}.

Bremsstrahlung from the same population of electrons, which flattens at 
non-relativistic energies, scales with the factor 
$\Sigma_i n_{i} Z_i^2 \kappa V/4\pi d^2$, 
where $\Sigma_i n_{i} Z_i^2$
indicates the sum of densities of all the ion species with charge $Z_i$. 
So from the combination of synchrotron emission and bremsstrahlung, 
and an estimate of  $\Sigma_i n_{i} Z_i^2$, we can estimate $\kappa$\ and $B$.

The radio spectrum of \casa\ has a spectral index of $\alpha = -0.78$ and based on a compilation of
radio data \citep[e.g.][]{Baars77} we estimate a radio flux density of
2522 Jy at 1 GHz for the epoch 2000. The spectral index indicates an electron 
energy spectrum with 
$q = 2.56$ at relativistic energies and flattening to 1.77 at 
non-relativistic energies, resulting in an X-ray photon index of -2.2.
This is at odds with the observed spectral index of -3.3, 
which means that 100 keV must have a different origin, taken here
to be bremsstrahlung from LHW accelerated electrons.
We adopt the LHW model, and try to obtain an upper limit on the emission from
an additional component  with a power law slope of -2.2  from the combined 
\sax\ and \osse\ data (Fig.~\ref{fig-osse}).

\medskip
\vbox{
\centerline{\psfig{figure=casa_cr_osse_pds.cps,width=8cm,angle=-90}}
\figcaption{
Best fit model that combines the LHW/\scff\ with the electron cosmic ray 
injection spectrum. Shown are the \sax-PDS (black) and \osse\ (red) spectrum of \casa.
\label{fig-osse}}}
\bigskip

The best fit normalization for such a power law spectrum is 
$(4.0\pm1.5)\times 10^{-7}$~ph s$^{-1}$keV$^{-1}$cm$^{-2}$ 
at 100~keV, with a 2$\sigma$ upper limit of 
$6.2\times 10^{-7}$~~ph\,s$^{-1}$keV$^{-1}$cm$^{-2}$.
We measured the electron spectrum normalization more directly by fitting a 
model of bremsstrahlung from a relativistically correct power law momentum 
spectrum, using the gaunt factors proposed by \cite{Haug97}.
The upper limit on the bremsstrahlung normalization obtained this way is
$\kappa \Sigma_i n_i Z_i^2 V/4\pi d^2 < 65$.
Soft X-ray observations imply $\Sigma_i n_i Z_i^2 \simeq 20$ 
\citep{Vink96,Willingale02b}, which, 
combined with the bremsstrahlung normalization, 
gives for the average magnetic field $B > 0.5$~mG.\footnote{This replaces
the value reported in \citet{Vink02}, 
in which we missed a factor $4\pi$\ in the synchrotron normalization 
\citep[eq.~\ref{eq-synchrotron}, c.f.][]{Ginzburg65}.} 
This is higher than the value derived by \citet{Atoyan00} from the \egret\
upper limit, especially taking into account that they
assumed a higher value for $\Sigma_i n_i Z_i^2$.
Note that the above lower limit is not a volume average, 
but based on the ratio $<\kappa B^{(q+1)/2}>/<\kappa \Sigma_i n_i Z_i^2 >$.
The derived lower limit is in good agreement with equipartition 
of magnetic and cosmic ray energy.

\begin{figure*}
\centerline{\psfig{figure=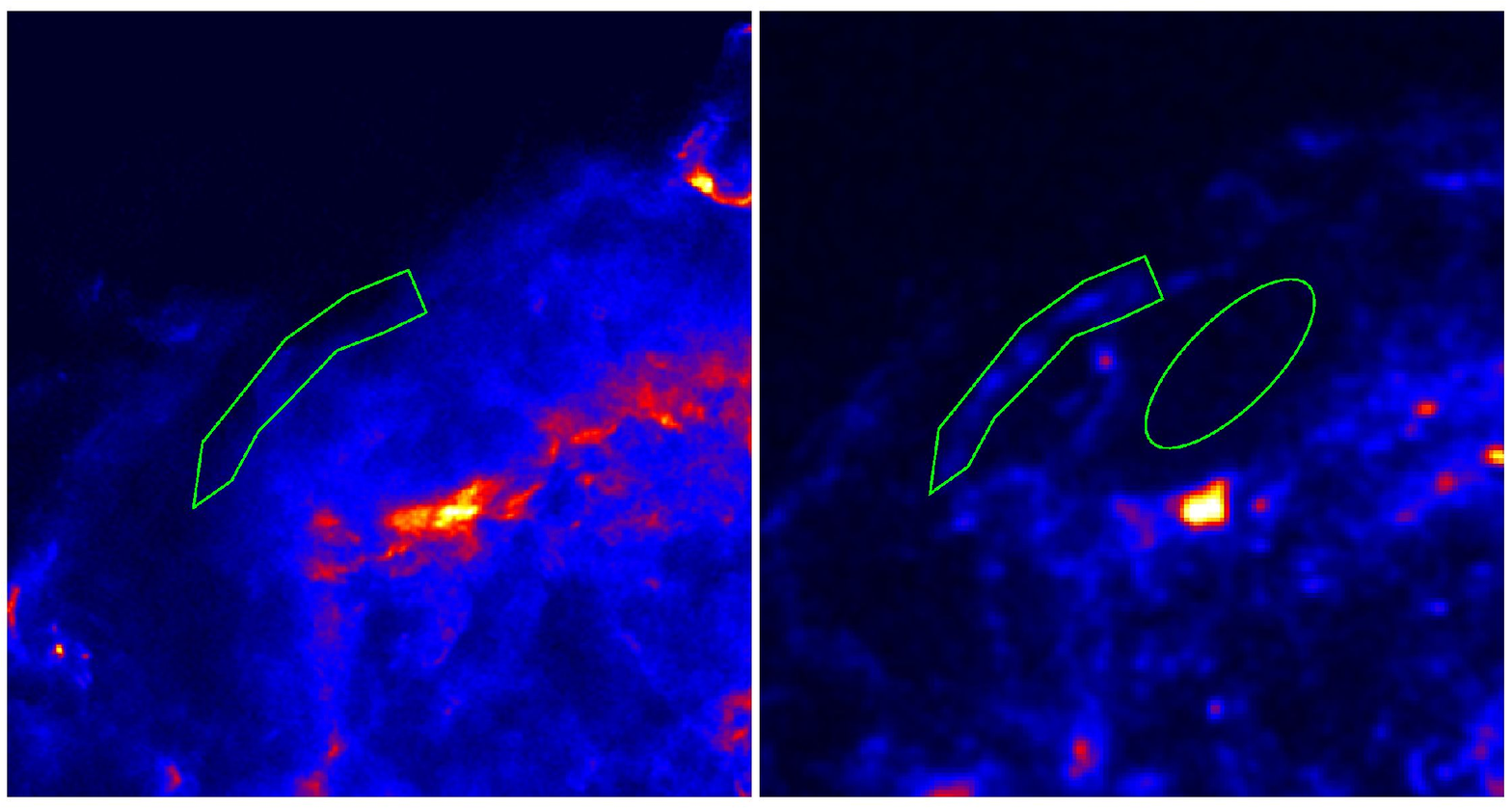,width=18.5cm}}
\figcaption{Detail of a VLA radio image (left) and \chandra-ACIS X-ray image from a compilation
of continuum dominated bands between 2.7 - 9 keV. The spectral extraction regions
are indicated in the \chandra\ image (NE rim and NE inside rim).\label{fig-image}}
\end{figure*}

\section{The magnetic field near the shock front}
The high average magnetic field in \casa\ is inconsistent with simple shock 
compression of the interstellar magnetic field, and indicates that the 
magnetic field inside the remnant
must have been enhanced by turbulence. A likely place for turbulence is the 
contact discontinuity, which may explain why the radio emission has a bright 
ring of emission,
as this is the region in the remnant where magnetic field amplification is 
important.
A similar argument was used by \citet{Anderson96}  to explain why radio knots
have the same radio spectral index as the surrounding diffuse emission;
the enhanced magnetic field lights up the background electron population.
One may wonder then, what the initial interstellar magnetic field was, 
before turbulent magnetic field amplification.

An estimate can be derived from the observation of the shock front of 
\casa\ by \chandra.
In the X-ray continuum band from roughly 4.5-6 keV \casa\ seems bound by a 
narrow rim of emission, which marks the onset of radio emission.
The narrowness of the rim indicates strong limb brightening of a layer,
which much have some special emission properties.
The fact that the radio and X-ray morphology are dissimilar as far as this 
rim is concerned, seems
to argue against a synchrotron origin for the X-ray emission, 
see e.g. \citet{Gotthelf01}.
But this is not true, if the width of the rim is a result 
of  the limited lifetime of the ultra-relativistic electrons, 
as soon as they are swept away from the shock front.
Further inside the shock the maximum energy of the electrons has so much 
decreased that no X-ray synchrotron emission is produced anymore.
This is not the case for the radio emission, 
caused by electrons with much lower energies, for which 
synchrotron and inverse Compton losses are negligible \citep[e.g.][]{Wright99}. 
This behavior can be seen in the models by \citet{Reynolds98}, 
in which the synchrotron emitting shell is thinner at higher frequencies. 
The steady increase in radio emission toward the bright ring of emission 
may be the effect of accumulation of relativistic electrons
swept away from the shock front, 
and a strong gradient of the magnetic field.

In this interpretation most of the electron populations, which reside outside
the X-ray rims, have a maximum energy cut-off well below 1~TeV, 
unless reacceleration is important.
However, radio spectral index measurements do not indicate any 
reacceleration of electrons by ejecta knots, which would likely lead 
to different indices of radio knots and diffuse emission \citep{Anderson96}.
Note that this argues also against the use for \casa\ of the simple 
maximum curved synchrotron model employed by \citet{Reynolds99}, 
which merely assumes that all electron populations
have the same energy cut-off, and the magnetic field is constant.

\medskip
\centerline{\psfig{figure=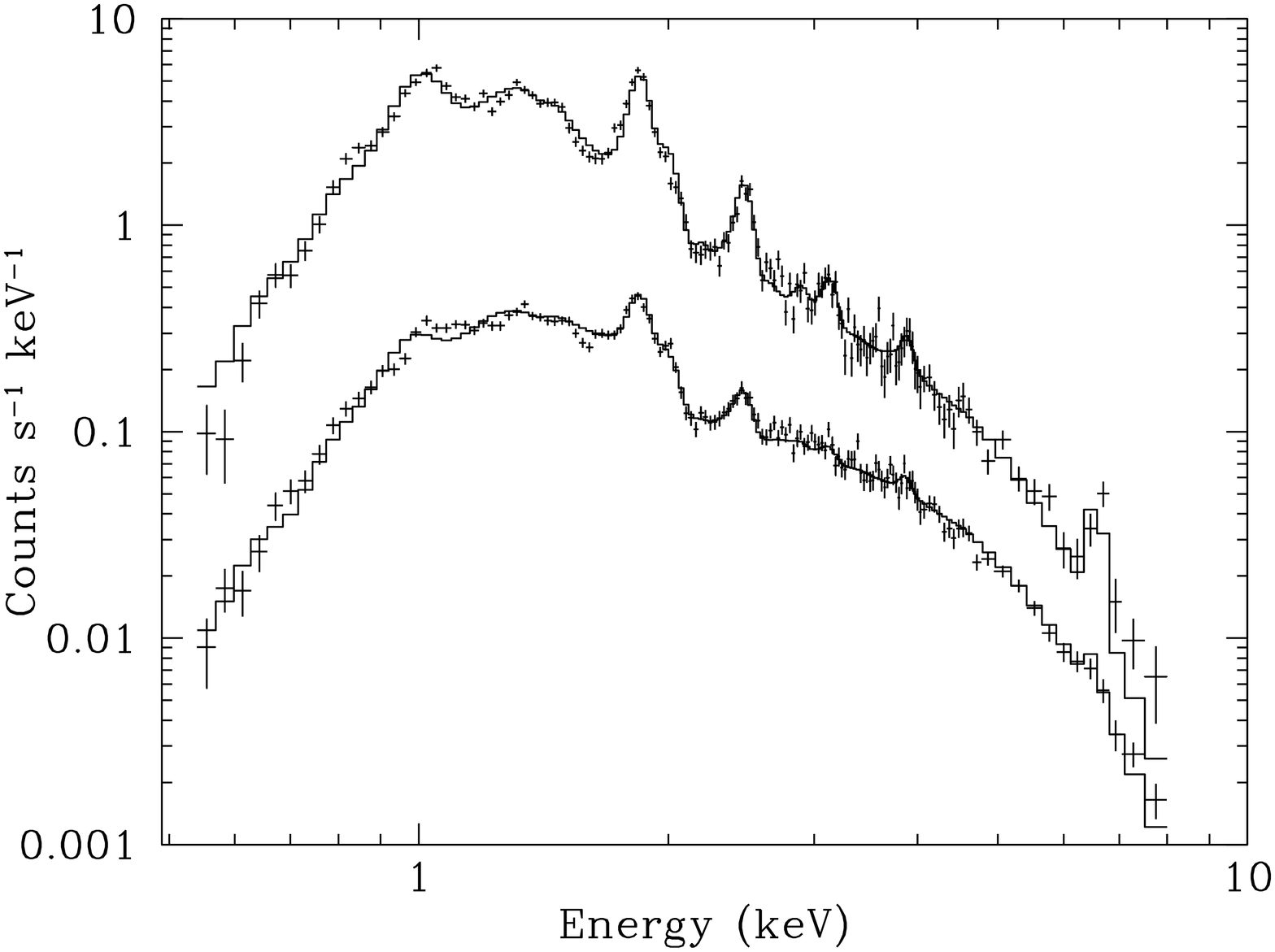,angle=-90,width=8cm}}
\figcaption{
\chandra\ ACIS-S3 X-ray spectra of the narrow rim of emission (lower
spectrum) and from emission just inside the rim (x10). Note the lower 
equivalent width of the line emission from the spectrum of the rim (cf. Table~\ref{tbl-chandra}).
\label{fig-spectra}}
\bigskip

The rim is much narrower than the $\sim$1/12th shock radius 
expected for plasma shocked with a factor 4 compression ratio.
That they are indeed consistent with X-ray synchrotron emission can be exemplified
by a comparison of the X-ray continuum flux and radio flux
of part of the rim in the northeast
(see Fig.~\ref{fig-image}, and Table~\ref{tbl-chandra}).
We used archival \chandra\ data to obtain a spectrum from this filament and
determined that the X-ray continuum flux between 4 and 6 keV is
$1.0\times10^{-12}$~erg~s$^{-1}$cm$^{-2}$ 
($2.1 \times 10^{-7}$~Jy at $1.2\times10^{18}$ Hz).
We normalized an archival VLA radio map (epoch 1987) to a flux of 2522~Jy 
and extracted a flux density from the same region. 
As the shock front may have moved between 1987 and 2001 by as much as 
5\arcsec, we extracted also a flux from the same region, 
but moved 5\arcsec\ toward the center of the remnant.
This yielded a flux density estimate between 3~Jy and 6~Jy at 1~GHz, 
corresponding to a broad band spectral index of $\alpha = -1.15$ 
(the factor two uncertainty in flux density is irrelevant over a frequency
range of 9 decades).
This is steeper than the radio spectral index of 0.78 and 
thus indicates that the X-ray flux density
is below the extrapolation of the radio spectrum; 
consistent with the interpretation that part of the X-ray emission 
is synchrotron emission.

Indeed, not all X-ray emission is synchrotron, 
as the narrow rim displays line emission from Mg, Si and S,
which should be accompanied by bremsstrahlung as well. 
The line emission from the rim is, however, 
weaker and the spectrum harder, than for the spectrum of a region just
inside of the rim (Table~\ref{tbl-chandra}).
For example fitting a thermal non equilibrium model with an additional power
law gives a spectral index of -2.2, 
whereas away from the rim the spectral index is 
unconstrained, but steeper than -4.
Moreover, it is hard to find another interpretation for the rim than
synchrotron emission, as only energy losses can explain the narrow features.
For a hot plasma the losses are small and the gradual ionization and possible 
equilibration of electrons and ions behind the shock would 
lead to a steady increase in intensity behind the shock, 
not in a sudden onset of emission, followed by a rapid decline.

Interestingly, identifying the rims with X-ray synchrotron 
emission from recently accelerated electrons makes this aspect of \casa\ very
similar to SN 1006, where the X-ray synchrotron emission is coming from
a similarly narrow rim of emission, as shown by archival \chandra\ data.
For SN 1006, however, the continuum emission from the rims are dominating
the total continuum emission, whereas for \casa\ the X-ray continuum morphology
shows that a substantial fraction of the X-ray continuum is associated with
the bright ejecta shell. In SN 1006 the synchrotron emission comes mainly from
the northeastern and southwestern limb, which is usually interpreted 
as caused by the large scale orientation of the magnetic field
\citep{Reynolds98}. 
For \casa, however, we do not see such axi-symmetry. 
This implies a much more randomly oriented magnetic field in
the surrounding medium.

Assuming that the limb brightened rim emits indeed X-ray synchrotron 
radiation, 
we can derive from the typical width of the rim
the magnetic field behind the shock front and the typical electron energy
responsible for the synchrotron radiation.
The electron energy may be associated with the typical exponential
cut-off energy of the electron spectrum, 
but the very nature of the losses
make it likely that a range of cut-off values exist. 
In an equilibrium situation synchrotron and inverse Compton losses will
steepen the spectrum by a $\Delta q = -1$ \citep{LongairV2}, 
but as soon as the electrons are
sufficiently far removed from the shock front the spectrum is more
likely to have an exponential cutoff. An exponential cut off is also more
likely, if the spectrum near the shock front is age limited instead of
loss limited \citep{Reynolds98}.

The rims have a typical width of  1.5\arcsec-4\arcsec\
and clearly stand out above the background and diffuse emission
(Fig.~\ref{fig-rims}).
We can estimate a typical loss time from the widths by combining
them with the plasma velocity away from the shock front.
The shock velocity in \casa\ has been estimated from proper motion 
studies using
Einstein and ROSAT HRI X-ray images, and is thought to be 
$v_s = 5200$~km~s$^{-1}$\ \citep{Vink98}.
The velocity of plasma away from the shock front, 
for a standard shock compression ratio of 4, is 0.25$v_s = 1300$~km~s$^{-1}$.
This value is consistent with X-ray Doppler shifts,
which imply a velocity relative to the shock front of
$\Delta v \sim 1400$~km/s \citep{Willingale02}.
For a distance of 3.4~kpc the measured widths of the rims correspond to
a typical loss time of 18 to 50~yr.

The energy history of an electron with initial energy $E_0$, 
suffering synchrotron losses, is:
\begin{equation}
E(t) = \frac{E_0}{1 + E_0 B^2 t/635}.
\end{equation}
After a time $\tau$ (eq.\ref{eq-losses}), the electron will have
an energy $\frac{1}{2}E_0$, whereas higher energy electrons, after the same
amount of time, will have energies $E^{\prime}(\tau) \leq E_0$.
The relation between the peak emitting frequency and 
electron energy is given by
\citep{Ginzburg65}:
\begin{equation}
\nu_{peak} = 1.8\times 10^{18} B_{\perp} E^2. \label{eq-syn}
\end{equation}
In reality we have to take into account a range of frequencies, as the
the synchrotron emission function is broad.

The rim is best visible in the X-ray continuum dominated band of 4-6 keV, 
and in some narrow bands in between line emission features. 
Taking 5~keV ($1.2\times10^{18}$~Hz) as the typical photon energy, 
we can combine eq. \ref{eq-syn} with eq. \ref{eq-losses} to obtain a typical
electron energy of $E = 65$~erg to $91$~erg 
and $B = 0.16$~mG to $0.08$~mG. 
The power slope of the Chandra spectrum near the shock is $-2.2$, which
indicates that the photons are emitted by electrons close to the break
in the electron distribution. In fact using the the \srcut-model 
(cf. \S \ref{sec-lhw}), we derive for $B = 0.1$~mG, $E_{break} \sim 25$~erg.
However, the spectrum is likely to be an integration of
spectra with different break energies. So close to
the shock front $E_{break}$\ is likely to be higher.

\medskip
\vbox{
\centerline{\psfig{figure=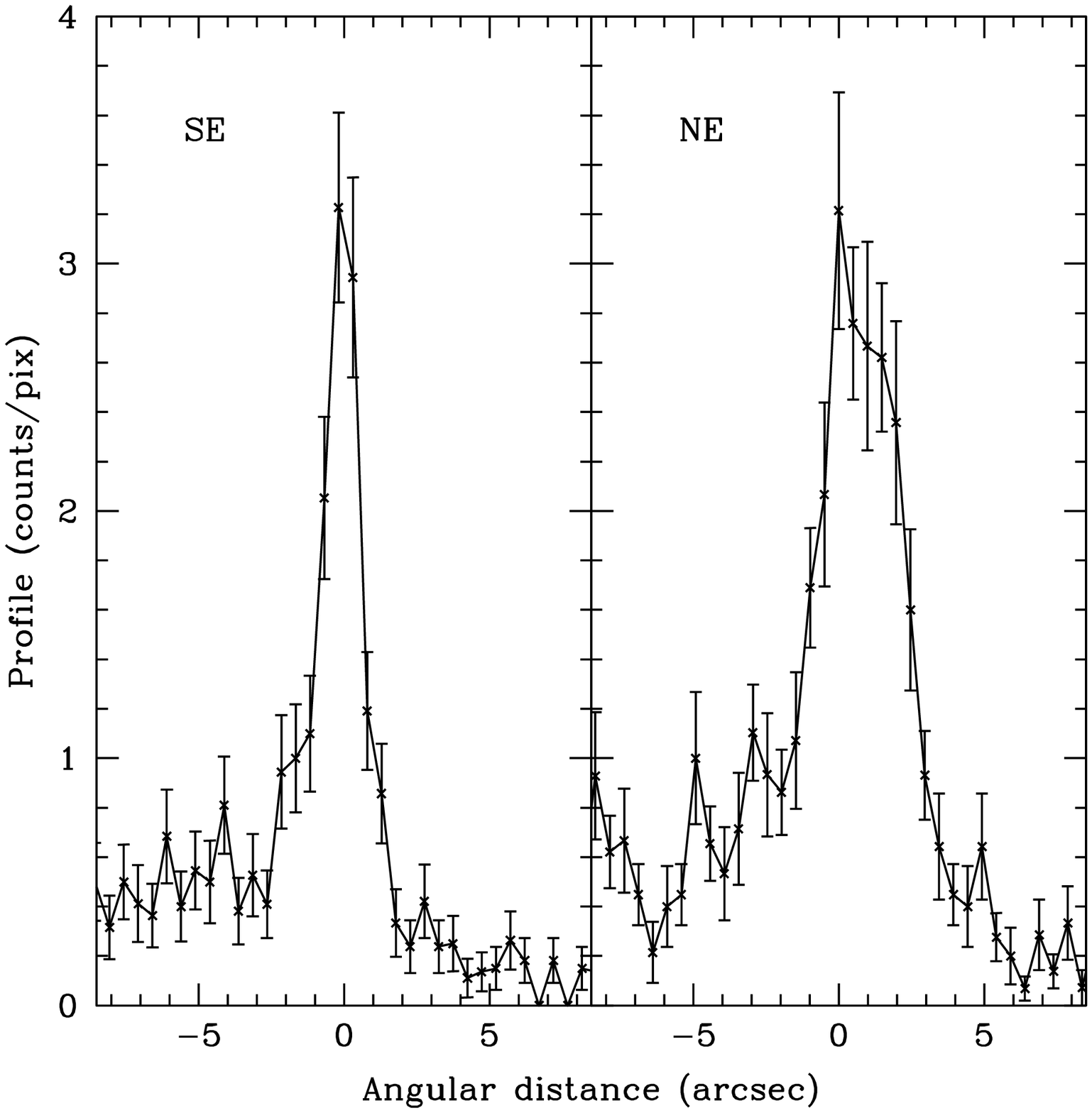,width=8cm}}
\figcaption{
Emission profiles of the rim in the southeast (left) and northeast. 
The profiles were extracted using 
strips of 20 pixels wide (9.8\arcsec) from a \chandra\ image, which was 
a combination of continuum dominated bands between 2.7 and 9 keV. 
\label{fig-rims}}}
\bigskip

In Fig.~\ref{fig-losses} we have illustrated our estimates of magnetic
field strength and electron energy graphically.
Apart from synchrotron losses we also have included inverse Compton 
losses. For this we have estimated the photon energy density from the cosmic
microwave background and the far infrared emission from \casa\ itself,
the latter being the most dominant factor. 
From the \iras\ infrared fluxes listed by \citet{Braun87} we estimate
a photon energy density of $u = 4.7\times10^{-12}$~erg\,cm$^{-3}$\ ,
corresponding to an equivalent magnetic field of $B = 1.1\times10^{-4}$~G.

The estimates of the magnetic field near the shock front give values
of roughly a fifth of the average magnetic field strength and indicate a 
preshock magnetic field of $\sim 2-4\times 10^{-5}$, a factor ten
higher than the canonical interstellar medium value.
This may either hint at a higher magnetic field surrounding \casa,
for instance because the wind of the Wolf-Rayet progenitor carried
a high magnetic field, or the presence of accelerated particles near
the shock front has given rise to magnetic field amplification due
to non-linear plasma wave growth.
The first hypothesis was suggested by \cite{Biermann93} in order
to explain the acceleration of cosmic ray by supernova remnants up to 
$10^{16}$~eV for which high magnetic fields are necessary.
The second hypothesis was suggested by \cite{Lucek00} for identical reasons.

\medskip
\vbox{
\centerline{\psfig{figure=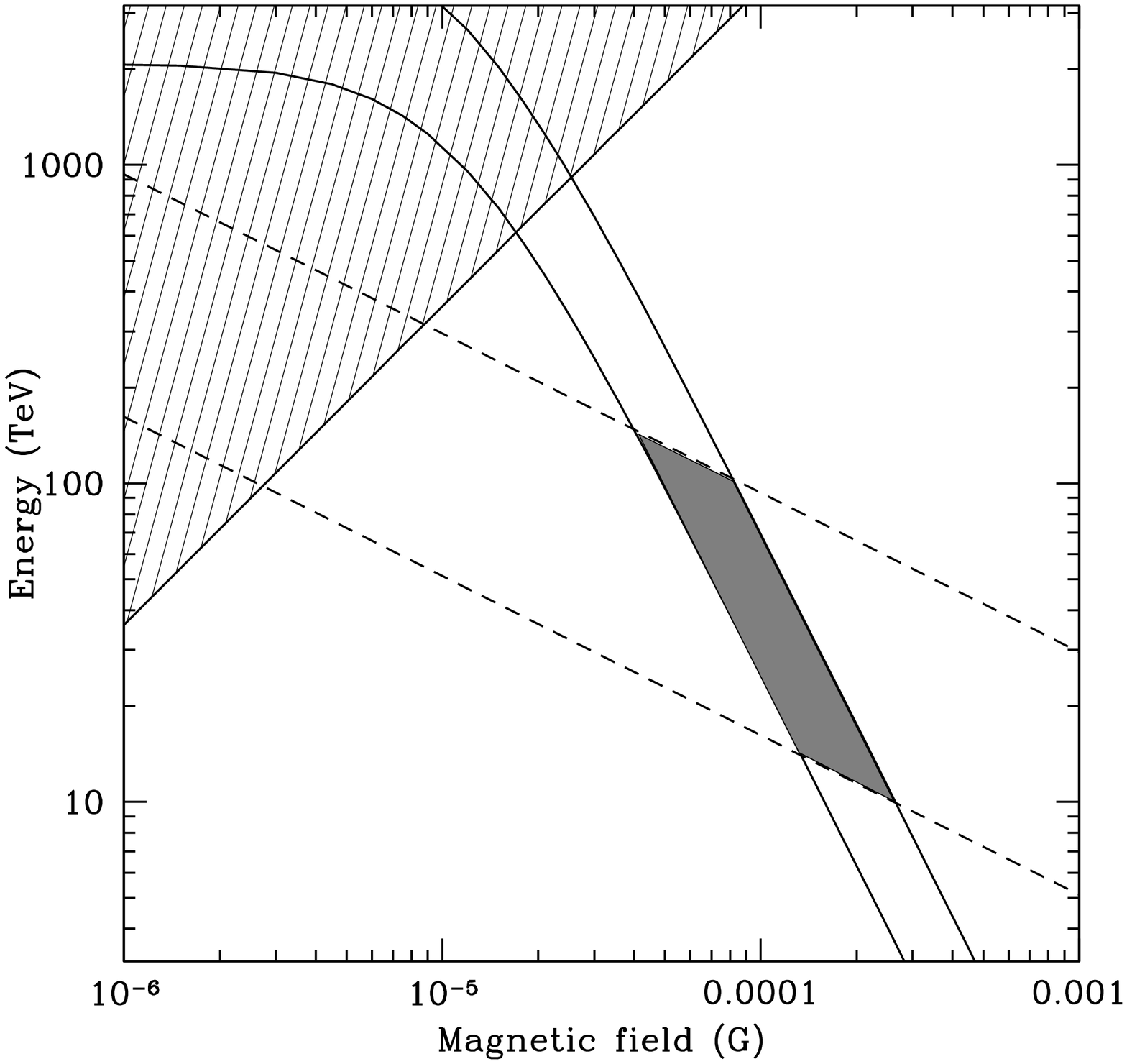,width=8cm}}
\figcaption{
The shaded area indicates the most likely values of $E_{max}$ and $B$ just 
behind the shock front. This area is formed by the two dashed lines,
which mark the electron energies 
that contribute to the continuum emission around 5~keV, allowing
for emission at 1/3rd and 10 times the 
peak frequency for synchrotron emission, and 
the solid lines, which are formed by
assuming the width of the rims is determined by radiative losses 
(decay times of 18 to 50 yr and a shock velocity of 5200 km/s are assumed).
The width of the rims should at least be larger than twice the gyro-radius.
This means that for a typical width of 3\arcsec, the hatched region is 
excluded.
\label{fig-losses}}
}
\bigskip

\section{The magnetic field strength and the nature of the TeV emission}

The measured magnetic field strengths have important consequences for
the gamma ray emission from \casa. For instance, our lower limit on
the average magnetic field strength implies that the gamma ray emission
from bremsstrahlung is almost an order of magnitude below the \egret\
upper limit.

It is also understood that for the high magnetic field in \casa, it is
unlikely that the detected TeV emission
is inverse Compton emission \citep{Aharonian01,Atoyan00}.
We have investigated this important issue by constructing
a two zone model. Zone 1 is responsible for the bulk of the radio
emission, such as the bright radio ring. It should have
$B > 0.5$~mG according to lower limit on the
magnetic field. As the electron population in this zone is old,
no X-ray synchrotron emission is produced. 
We adopt a somewhat arbitrary maximum electron energy of 0.1~TeV, 
but in reality a range of values exist inside the remnant
The inverse Compton emission from this zone is negligible. 

Zone 2 is responsible for the X-ray synchrotron emission
from the rim and has $B \sim 0.08 - 0.16$~mG and $E_{max} \sim 40 - 60$~TeV.
Based on our flux measurements, and assuming spherical symmetry, we estimate
that the X-ray synchrotron emission should be less than one third of the
X-ray emission around 5~keV, but as we focused on a particular bright
part of the rim, 1/10th to 1/20th seem more likely values.
Moreover, for  $B = 0.08$~mG and an X-ray synchrotron flux contribution of
1/3rd, the non-thermal bremsstrahlung flux at 100 keV is above the measured 
upper limit. This criterion is relaxed for higher magnetic
fields near the shock front.

\medskip
\centerline{\psfig{figure=casa_broadband_ir72_e55tev_b1e-4g.ps,angle=-90,width=8cm}}
\figcaption{
A simplified two zone model of the broad
band radiation from the cosmic ray electron population,
i.e. synchrotron radiation (dotted line), bremsstrahlung
(dashed-dotted), and inverse Compton emission( dashed line).
The solid line gives the total emission from both zones. 
The parameters near the shock front (zone 2) were chosen to be
$B= 1.0\times10^{-4}$~G, $E_{max} = 55$~TeV. $B=0.5$~mG for zone 1.
The data points give the flux at 1 GHz, measurements by 
\citet{Mezger86},
X-ray continuum emission (\asca\ and \sax), \comptel\ \citep{Strong00}, 
the \egret\ upper limit \citep{Esposito96}, and \hegra\ \citep{Aharonian01}.
\label{fig-broadband}}
\bigskip

This two zone model is different from the two zone model proposed
by \citet{Atoyan00}, for which the two zones relate to the
radio knots and diffuse radio emission. We adopt here the view
of \citet{Anderson96}, that there are no intrinsic differences between
the electron cosmic ray populations of the radio knots and the diffuse emission,
except that in the radio knots the magnetic field is stronger.
Note that the electron cosmic ray density is higher in zone 2. Zone 1 is brighter
at radio wavelength due to the stronger magnetic field.

In order to estimate the inverse Compton emission we need to estimate the
photon density caused by \casa's far infrared emission. 
\iras\ flux measurements \citep{Braun87} indicate a total photon flux of 
$F_{ph} =3.9\times10^{5}$~{ph s$^{-1}$ cm$^{-2}$},
centered around $10^{13}$~Hz.
The far infrared emission is associated with ejecta \citep{Lagage96}.
Approximating the geometry by a sphere with  radius similar to the ejecta, 
$r = 1.8$~pc, 
and assuming that half the number of photons spend an average time of 
$r/c$ inside the sphere, 
we obtain for the photon density\footnote{The other half of the photons
escape directly, but all photons will have to pass through zone 2.}:
\begin{equation}
n_{ph} = \frac{3}{2} \frac{d^2}{r^2} \frac{F_{ph}}{c} = 72 {~\rm cm^{-3}},
\end{equation} 
where $d = 3.4$~kpc is the distance to \casa.
A similar reasoning gives a somewhat
lower photon density for zone 2, but the photon field is much more
anisotropic, on average coming from the direction of the remnant's center.
For the spherical symmetry assumed, the anisotropy has
no influence on the total inverse Compton emission, but as the maximum
upscattering occurs for head on collisions, most energetic inverse Compton emission
from zone 2 will be directed toward the center of the remnant, giving rise to
a centrally brightened morphology, despite the fact that
according to our model the dominant inverse Compton emission
is coming from a narrow shell near the shock front.
In contrast, bremsstrahlung from the same electron population will show
a ring-like morphology.
Unfortunately, these interesting differences in morphology occur on the
scale of arcminutes, which is currently beyond the reach of gamma-ray
instruments.

Our two zone model is illustrated in  Fig.~\ref{fig-broadband}, 
employing typical values for $B$ and $E_{break}$ near the shock front.
The inverse Compton emission is completely dominated by the electrons in 
zone 2. 
In  Fig.~\ref{fig-broadband} inverse Compton emission 
is well below the TeV flux by \hegra. 
It is therefore likely that the TeV emission is dominated by pion decay. 
However, we cannot yet make a definitive statement regarding
the inverse Compton emission. 
For lower values of the magnetic field zone 2 is also the dominant
source of bremsstrahlung by relativistic electrons.
If the magnetic field is at the low
end of our inferred magnetic field, $B= 0.08$~mG, and 
the far infrared photon density is larger than 150~cm$^{-3}$, 
the \hegra\ detection, and the \sax/\osse\ upper limit on
the non-thermal bremsstrahlung are still consistent
with inverse Compton emission.
The basic uncertainty in the photon density stems from the uncertain
geometry, which for zone 2 may also include anisotropic photon emission.
The combination of a low magnetic field and a rather high infrared
photon density seems unlikely, but is not impossible.
Even lower magnetic fields in zone 2 are unlikely, 
as  the non-thermal bremsstrahlung would 
exceed the upper limit reported in \S 3.

\section{Conclusion}
We have provided estimates of the average magnetic field
and the magnetic field near the shock front. 
The lower limit to the average magnetic field is based on the assumption
that the electron spectrum follows a 
power law distribution in momentum, or is steeper than that at energies between
100-1000 keV.

The magnetic field strength near the shock front, was measured
from the width of the narrow rim surrounding \casa\ in X-ray continuum
images, assuming that the emission mechanism is synchrotron emission, and
the width is determined by synchrotron losses.
The contribution of the X-ray synchrotron emission to the X-ray continuum
flux is less than one third.
This is supported by the X-ray morphology, 
which shows that even around 10~keV a substantial part of the
continuum is associated with the bright ejecta shell \citep{Bleeker01}.

This justifies taking into consideration alternative models than
synchrotron radiation for the hard X-ray emission of \casa.
This is was done in \S 2, where we tested the lower hybrid wave
model.
The bremsstrahlung spectrum predicted for electrons
accelerated by lower hybrid waves has a similar spectral
slope as is observed. The maximum energy to which the electrons are
accelerated by this mechanism is $\sim$95~keV, according to fits
to the \sax\ data.

The estimated lower limit on the average magnetic field is 0.5~mG, whereas
we find $\sim 0.1$~mG near the shock front. So even near the shock front
the magnetic field strength is an order of magnitude above what is expected
from shock compression of the canonical interstellar magnetic field of 
3~$\mu$G. This can be explained by either a higher magnetic field strength
in the medium surrounding \casa, 
possibly as the result of the stellar
wind of the progenitor \citep{Biermann93},
or it means that the magnetic field is being rapidly enhanced near the
shock front \citep{Lucek00}.

Electron energies close to the shock front may reach $\sim$50~TeV or higher,
but in contrast to  \citet{Reynolds99},
we suggest that most of the cosmic ray electrons
responsible for the bright radio ring are in an environment with a high magnetic field,
where due to synchrotron losses the the cut-off energy is well below the energy 
where X-ray synchrotron emission can be expected.

The high average magnetic field makes it likely that TeV emission from
\casa\ is dominated by pion decay \citep{Atoyan00}, but there is
some uncertainty, as in our two zone model of \casa\ it is still possible
to obtain a high inverse Compton flux by assuming that the magnetic
field is at the low end of what we have inferred from the X-ray rims, $B=0.08$~mG,
and using a higher far infrared photon density of 150~cm$^{-3}$.

Future mission will improve on those results. For instance, \glast\ will
probably be able to detect bremsstrahlung in the GeV range,
which will provide a much more direct way of estimating the average
magnetic field of \casa.

As for the hard X-ray emission.
\integral\ will not greatly improve the current measurement of
the hard X-ray continuum emission, but by accurately measuring
the \tiff\ related line emission, it will constrain that component
of the emission around 80~keV, and thus indirectly improve our understanding
of the continuum emission around 100~keV.

One of the consequences of our analysis is that the gamma-ray emission
will be mostly associated with the outer rim of \casa, 
not with the bright X-ray and radio shell.
Unfortunately, the spatial resolution required to measure this will
not be available in the near future.

Finally, as an interesting by product of our analysis, we mention that
the new modeling of the hard X-ray emission with the lower hybrid model
yields revised values of the \scff\
line emission at 68~keV and 78~keV, which correspond to
a 50\% higher mass of \tiff\ synthesized in the explosion compared to
what was reported by \citet{Vink01}, i.e. $M(^{44}{\rm Ti}) = 1.8\times10^{-4}$~\msun.
This is closer to the latest \comptel\ measurements \citep{Schoenfelder00}.

\acknowledgements
JV was supported for this work by the NASA
through Chandra Postdoctoral Fellowship Award Number PF0-10011
issued by the Chandra X-ray Observatory Center (CXO), which is operated by the
Smithonian Astrophysical Observatory (SAO) for and on behalf of NASA under contract
NAS8-39073.
JML was supported by basic research funds of the Office of Naval Research.
We acknowledge the use of the HEASARC (Goddard Space Flight Center), 
CXO (SAO), and NCSA Astronomy Digital Image Library archival facilities.

\appendix
\section{Appendix}
In this appendix we want to demonstrate how free parameters in the fit of
the model continuum (the maximum accelerated electron energy and the fraction of
ambient electrons that are accelerated) are in principle determined by the plasma
physics parameters of Cas A. The maximum electron energy is related to the
maximum electric field that a lower hybrid wave can develop and the time that an
electron can spend in this field, governed by the thickness of the shock precursor.
The fraction of electrons that may be accelerated is expected to be determined by
a marginal stability criterion, i.e. the electrons will be accelerated until the
heating of the ambient plasma produced by Coulomb colisions of the accelerated
electrons violates the conditions necessary for lower hybrid waves to exist, i.e. that
the electron gyroradius be less than the wavelength divided by $2\pi$.

The lower hybrid waves are generated by cross field ion motions produced by
shocks. A fraction of preshock ions are reflected back upstream
ahead of the shock with a velocity relative to the preshock
plasma of $\sim 2v_s$, where $v_s$
is the shock velocity. Ions reflected initially back along the shock
velocity vector at a perpendicular shock travel a distance $0.68v_s/\Omega _i$
before returning \citep{gedalin96}. Here $\Omega _i$ is the ion gyrofrequency.
The maximum distance ahead of the shock
that reflected ions can reach is $d=1.37v_s/\Omega _i$. This obtains for ions
reflected from the shock at an angle $30^{\circ}$ to the shock
front, so that $2v_s\sin 30^{\circ} =v_s$.

Consider a shock front
at an angle $\theta$ to the local magnetic field direction. In
order for the lower-hybrid waves generated ahead of the shock by
the reflected ion precursor to grow to large amplitudes we require
\citep[][i.e. the wave group velocity away from the shock is equal to the shock
velocity, so that the time available for wave growth is essentially unlimited on
the timescale of these processes]{Laming01b}
\begin{equation}
{\partial\omega\over\partial k_{\perp}}= {v_s\over \cos\theta}
={\alpha\omega\over k{\perp}}
\end{equation}
where $\alpha = 1/\left(2\cos\beta\cos\theta\right)$ and $\beta$ is the angle
between $\partial\omega /\partial \vec{k}$ and the reflected ion
bulk velocity vector $\vec{U}$. A minimum requirement for wave
growth is $-1 < \alpha < -1/2$. Since it is ions {\em returning} to the shock
that can generate the necessary waves, waves generated at
the earliest possible period in the reflected ion orbit, i.e. lowest value of
$\left|\cos\beta\right|$, when
$\alpha\sim -1$, (for $\cos\theta\sim 1$) will dominate the growth.

The maximum electric field that the lower hybrid wave may develop
before ion trapping and heating set is given by
\citet[in mks units]{karney78};
\begin{equation}
E=B\left(\Omega _i\over\omega\right)^{1/3}{\omega\over 4
k_{\perp}} \simeq B\left(\Omega
_i\over\omega\right)^{1/3}{v_s\over 4\cos\theta}.
\end{equation}
Taking the wave frequency $\omega =\Omega _{LH}=\sqrt{\Omega _i\Omega _e}$, and the
component of the electric field directed along the magnetic field direction to be
$E\omega _{pi}/\omega _{pe}$
gives $a\simeq\Omega _{LH}v_s/16\cos\theta$ for electron acceleration along the
magnetic field direction. $\Omega _{LH}$ is the geometric mean of the
ion and electron gyrofrequencies and is known as the lower hybrid frequency.

We treat the case of a nonrelativistic shock which may accelerate
electrons to relativistic energies. We work in the rest frame of
the preshock medium. In this frame an electron is accelerated by
the component of the electric field directed along the magnetic
field, until it outruns the reflected ion shock precursor a
distance $d$ from the shock front. Assuming acceleration begins
with the electron at the shock front moving along $\vec{B}$ with
initial velocity $v_0$, it will continue for a time $t$ given by
\begin{equation}
\left\{{c^2\over
a}\sqrt{1+{\left(at+v_0/\sqrt{1-v_0^2/c^2}\right)^2\over c^2}}
-{c^2\over a\sqrt{1-v_0^2/c^2}}\right\}\sin\theta = d+v_st.
\end{equation}
which gives a quadratic equation for $t$:
\begin{equation}
\label{quadratic}
t^2\left(c^2\sin ^2\theta -v_s^2\right) +t\left\{{2c^2\over
a\sqrt{1-v_0^2/c^2}}\left(v_0\sin ^2\theta
-v_s\sin\theta\right)-2dv_s\right\}-d^2-{2dc^2\sin\theta\over a\sqrt{1-v_0^2/c^2}}=0
\end{equation}
We require $\sin\theta \ge v_s/v_0$ in order that the electron stays
ahead of the shock front, and maximum acceleration time will be
realized for $\sin\theta = v_s/v_0$. In this case $\cos\theta\simeq 1$ and
the term in $t$ in
equation \ref{quadratic} simplifies to $-2dv_st$ and
\begin{equation}
t={d\over v_s\left(c^2/v_0^2-1\right)}\left\{1+\sqrt{1+\left(c^2/v_0^2-1\right)
\left(1+{2c^2\over adv_0v_s\sqrt{1-v_0^2/c^2}}\right)}\right\}.
\end{equation}
For $v_0<<c$,
$at=\sqrt{2adv_0/v_s}$ and for $v_0\sim c$, $at=2ad/vs\left(c^2/v_0^2-1\right)$.
Since $ad\propto v_s^2$ it is easy to see that the highest electron energies
are most efficiently reached by fast shocks in hot plasma, i.e. for the highest
plausible values of $v_s$ and $v_0$ in Cas A. At $\sin\theta= v_s/v_0$, three to four
shock interactions are required to accelerate electrons to the observed maximum energy
of $\sim 95$ keV, increasing to about ten shock interactions if $\sin\theta=1.2v_s/v_0$,
for $v_s=3000$ km s$^{-1}$ and $v_0=3\times 10^9$ cm s$^{-1}$ (corresponding to an
electron temperature of $4\times 10^7$ K). Even faster shock require fewer interactions.
The maximum energy in the accelerated
electron distribution will be determined by the number and obliquity of shock
interactions experienced by these electrons during the evolution of Cas A, and so for the
time being remains uncertain. However the maximum energy inferred from Fig. 1 is
certainly plausible in these terms. In principle,
given enough shock interactions, relativistic electrons can be produced. If a means
of transporting these electrons to the forward shock were available, this might
have a bearing on providing an injection mechanism for further electron acceleration
by a Fermi mechanism.

The normalization of the non-thermal bremsstrahlung spectrum is
determined by the fraction of plasma electrons that are
accelerated in this way, and this will be determined ultimately by
a marginal stability criterion. The instability requires the
electron gyroradius to be less than the wavelength divided by
$2\pi$, and once a sufficient number of electrons are accelerated to heat the
ambient plasma to temperatures higher than this, the instability
will shut down. The ambient plasma loses energy through radiation,
adiabatic expansion and possibly conduction. In \citet{Laming01b} the
heating rate due to the observed accelerated electrons is estimated to be
$4\times 10^{-18}n_en_e^{\prime}$ ergs cm$^{-3}$s$^{-1}$ for ambient and
accelerated electron densities $n_e$ and $n_e^{\prime}$. The electron heating
rate is $dT/dt = 0.019n_e^{\prime}$ K s$^{-1}$, and equating this to the
cooling rate by adiabatic expansion of the Cas A remnant yields an accelerated
electron fraction $n_e^{\prime}/n_e \sim 0.02$. The inclusion of ionization and
radiative power losses and relaxing the steady-state assumption inherent in this
estimate will allow higher accelerated electron fractions.

In \citet{Laming01b} an accelerated electron fraction of 4\% was found to give
a good match to the $\sim 50$ ksec of BeppoSAX MECS and PDS data on Cas A then available.
The estimate herein using only the $\sim 500$ ksec PDS data is higher at around
10\% (see Fig. 1), but a range of plasma temperatures exists in Cas A \citep[c.f.][]{Vink96,Willingale02}, 
and only the hottest plasma with a temperature of $\sim 4 \times 10^7$~K contributes
to the PDS bandpass. So with respect to all the hot plasma, an accelerated electron fraction of $\sim$4\% is
consistent with the data, and preferred on theoretical grounds.
Simulations of the evolution of the electron temperature following electron heating
indicate that such an accelerated electron density heats plasma in the 
location of
the contact discontinuity to temperatures in the range $4-5\times 10^7$ K.
These temperatures are basically constant over the past $\sim 30$ years over 
which X-ray observations have been made. 
In this respect simulations of Cas A expanding into
an $\rho\propto 1/r^2$ circumstellar medium appear to represent the 
physics much better than the simulations using a uniform density circumstellar
medium in \citet{Laming01b}.

\begin{deluxetable}{lccccccccc}
\tabletypesize{\scriptsize}
\tablecaption{Summary of model fits to the \sax-PDS spectrum.
\label{tbl-lhw}}
\tablewidth{0pt}
\tablehead{& LHW model & {\it srcut} & {\it sresc} & power law}
\startdata
Norm continuum\tablenotemark{a}	& $0.058 \pm 0.002$ & $2522$   	  	    &  $2522$           & $2.37\pm0.2$\\
power law index 		&			& 		    & 	                & -$3.32\pm0.05$\\
$E_{max}$/$E_{cut off}$ (keV)	& $95 \pm  5$       & $1.21\pm0.01$         &  $3.42 \pm 0.02$  &\\
Norm thermal emission\tablenotemark{b}	& $0.62 \pm 0.02$ & $7.6 \pm 3.3$    &  $0.0 (<2.5)$& $0.0 (< 17)$\\
\scff\ line flux ($10^{-5}$ \fluxunit) & $3.2 \pm 0.3$ 	& $2.7 \pm 0.4$     &  $2.9 \pm 0.4$ &  $2.1\pm0.4$\\
Ta K$\alpha$ flux ($10^{-5}$ \fluxunit)& $2.4 \pm 0.7$ &     $2.6 \pm 0.8$  &  $2.8 \pm 0.8$ &  $1.4\pm0.$\\
$\chi^2$/d.o.f          	& 273.1/251         & 273.0/252             & 289.3/252     & 267.5/251\\
\enddata
\tablenotetext{a}{
The normalization for the LWH model is defined as $10^{12}$ \EM\ cm$^{-5}$.
The normalization for the synchrotron models is fixed to the radio flux 
density at 1~GHz for the epoch 2000.0, and radio spectral index is 0.775. 
These values are based on a compilation of radio data, and includes
the effect of the secular flux decrease of \casa\ \citep{Baars77}.
The normalization for the power law model is ph/s/keV at 1~keV.}
\tablenotetext{b}{The normalization is defined as  $10^{12}$ \EM.
The temperature was fixed to $4 \times 10^7$~K.
Note that this definition assumes solar abundances, which is
not applicable for \casa, so care should be taken before using
it to calculate electron densities.}
\end{deluxetable}

\begin{deluxetable}{lcc}
\tabletypesize{\scriptsize}
\tablecaption{Best fit parameters for spectra from the northeastern rim.\label{tbl-chandra}}
\tablewidth{0pt}
\tablehead{& \colhead{NE Rim} & \colhead{Inside NE Rim}}
\startdata
$kT$   (keV) & $3.7\pm0.7$ & $3.0\pm0.4$\\
$n_{\rm e} t$ ($10^{10}$~cm$^{-3}$ s$^{-1}$) & $3.3\pm 0.5$\tablenotemark{a} & $5.2\pm0.6$\\
Ne & 	\multicolumn{2}{c}{$0.4^{+1.0}_{-0.4}$} \\
Mg & 	\multicolumn{2}{c}{$0.9 \pm 0.4$} \\
Si & 	\multicolumn{2}{c}{$4.5 \pm 0.6$} \\
S  & 	\multicolumn{2}{c}{$4.3 \pm 0.7$} \\
Ar & 	\multicolumn{2}{c}{$4.4 \pm 1.5$} \\
Ca & 	\multicolumn{2}{c}{$5.4 \pm 2.5$} \\
Fe & 	\multicolumn{2}{c}{$2.2 \pm 0.4$} \\
EM \tablenotemark{c} ($10^{10}$~cm$^{-5}$) & $3.0\pm0.4$ & $7.5 \pm 1.0$\\
PL index & $-2.24 \pm 0.05$& $-4.5 (< -4.3)$\tablenotemark{d} \\
PL norm ($10^{-3}$~s$^{-1}$cm$^{-2}$ \tablenotemark{e} )& $2.5 \pm 0.2$& $3.4 \pm 0.6$\\
PL contribution 4 - 6 keV & 84\% & 11\%\\
$N_{\rm H}$ ($10^{22}$~cm$^{-2}$) & \multicolumn{2}{c}{$1.08 \pm 0.03$}\\
$\chi^2/\nu$ & 261/161 & 363/153\\
\enddata
\tablecomments{
The spectral model consisted of \spex's  non-equilibrium ionization model (v1.10),
and a power law (PL) continuum, in order to approximate the synchrotron radiation.
The column density and abundances were first determined for 
the spectrum inside the NE Rim, and subsequently applied to the rim spectrum.
Following \citet{Vink96} we set He and N abundances to 10 times solar \citep{Anders89}.
The poorly constrained O abundance was coupled to Si; Ni
was coupled to Fe. Errors correspond to 95\% confidence intervals.
}
\tablenotetext{a}{Note that 
this value is roughly consistent with the synchrotron loss times,
if the electron density is $n_{\rm e} \sim 30$~cm$^{-3}$.}
\tablenotetext{b}{The He and N abundances were assumed to be 10 times solar, following
\citet{Vink96}.}
\tablenotetext{c}{The emission measure (EM) is defined as \EM.}
\tablenotetext{d}{The power law slope was poorly constrained, 
but tended to be steep, we therefore fixed it to a value of 4.5. 
The lower limit corresponds to $\Delta \chi^2 = 4$.}
\tablenotetext{e}{At 1~keV.}
\end{deluxetable}

\end{document}